\documentclass[npg]{copernicus}
\usepackage{graphicx}
\usepackage{amsmath}
\usepackage{amssymb}
\usepackage{times}                   

\newcommand{\beq}{\begin{equation}}
\newcommand{\eeq}{\end{equation}}
\newcommand{\beqa}{\begin{align*}}
\newcommand{\eeqa}{\end{align*}}
\newcommand{\mean}[1]{{\langle #1 \rangle}}

\begin{document}

\title{Correlated earthquakes in a self-organized model}
\runningtitle{Correlated earthquakes in a self-organized model}
\runningauthor{M. Baiesi}

\date{\today}

\author{Marco Baiesi}
\affil{Instituut voor Theoretische Fysica, K.U.Leuven, B-3001, Belgium}

\maketitle

\abstract{
Motivated by the fact that empirical time series of earthquakes
exhibit long-range correlations in space and time and
the Gutenberg-Richter distribution of magnitudes, we propose a simple
fault model that can account for these types of scale-invariance.
It is an avalanching process that displays power-laws in the event sizes, 
in the epicenter distances as well as in the waiting-time distributions,
and also aftershock rates obeying a generalized Omori law.  
We thus confirm that there is a relation between temporal and
spatial clustering of the activity in this kind of models.
The fluctuating boundaries of possible slipping areas show that
the size of the largest possible earthquake is not always maximal, and the
average correlation length is a fraction of the system size.
This suggests that there is a concrete alternative to 
the extreme interpretation of self-organized criticality as
a process in which every small event can cascade to an arbitrary large one:
the new picture includes fluctuating domains of coherent stress 
field as part of the global self-organization. Moreover,
this picture can be more easily compared with other scenarios discussing
fluctuating correlations lengths in seismicity.}


\section{Introduction}\label{sec:intro}
At the moment there is not a comprehensive explanation of the mechanisms
giving rise to the complex phenomenology of earthquakes.
The magnitude of each earthquake is characterized by the Gutenberg-Richter 
(GR) law~\citep{gutenberg}, which is in fact a scale-invariant 
distribution of energy release. 
Earthquakes are also long-range correlated with each other.
It is indeed known that events are clustered
in space and time~\citep{turcotte_book,scholz_book} 
and take place in complex fault patterns~\citep{bonnet01:faults}. 
The Omori law of aftershocks rate~\citep{utsu95:_omori}
is an example of the temporal clustering of earthquakes, 
with a decay given by a scale-invariant law. 
The phenomenology of the distance between subsequent epicenters is
also characterized by power-law 
distributions~\citep{davidsen-paczuski05:jumps,corral06:_jumps}. 
Moreover, the values of magnitudes, waiting times and locations of 
earthquakes are part of a single scaling 
picture~\citep{bak02:_unified,corral03:_unified,corral04:_clustering,corral05:_bayesian}.
Other examples are given by~\citet{mega03} and \citet{davidsen06:_record}.
Since seismicity is one of the most outstanding examples of a class of 
phenomena involving a wide range of energetic, spatial, and 
and temporal scales, it is expected that its modeling is problematic.

It is possible to build models based upon the phenomenology of earthquakes.
For example, aftershock-sequence models require
an assumed law of off-spring generation per 
event~\citep{ogata88:_ETAS,helmstetter02:ETAS,turcotte07:_BASS,
lippiello07:ETAS}. 
These models can yield realistic time-series,
but by construction they use rather than explain laws like the GR one.

The scale-invariant distribution of earthquake sizes is reproduced
by processes based on avalanches of stress redistribution,
following the idea that there is self-organized criticality 
(SOC)~\citep{bak_book,sornette_book}.
The precursor of this concept in geophysics has been the slider-block model
by~\citet{burridgeknopoff}. 
It is evident from many models that the mechanism of avalanches of
relaxations robustly leads to size-frequency power-laws. 
This behavior emerges from the 
collective organization of units that cooperate with
very nonlinear rules, redistributing stress
and typically dissipating it from open boundaries.

However, it has become also clear during the last years that the simplest
SOC models cannot reproduce other important features of critical phenomena,
usually involving correlations between events.
{ Models incorporating correlated events~\citep{olami92:_OFC,
hainzl99:SOC,hainzl00:SOC,
hergarten02:_OFC_aftershock,zoeller05:creeping,
huang98:_BTW-hier,lippiello05:_SOC_memory,
BM_05_soc,lippiello06:on-off,abaimov07:_recurr}
are a minority within the literature on SOC.
These few scattered results unfortunately have not constituted a 
large enough body for appropriately raising the issue of temporal 
organization to the attention of the scientific community.}

In this paper we show that earthquakes phenomenology
can guide us to build self-organized models with the appropriate features.
In particular, we stress the importance of clustering events in space and time,
an aspect leading us to develop a fault model that displays 
a full spectrum of power-law statistics (GR law, Omori law, waiting times and
epicenter distances with broad distributions), not observed in previous models. 
Hence, the very basic idea of SOC is in fact achievable. In particular,
the process self-organizes the epicenter locations, clustering them rather
than spreading them randomly in space, as it is frequently imposed in 
other simple models.

{ A novel feature distinguishing the model we propose from previous ones is
is the possibility to infer maximal areas of events from its 
configuration.}
It turns out that 
this model does not conform to the common picture associated
with SOC in geophysics \citep{nature-debate,geller97:_science}.
The idea is that every tremor can in principle
cascade in a large event, depending on minor details of the stress field.
It is possible that the paradigm of sandpiles has been much influential
in the consolidation of this view.
Up to date, this interpretation has been a speculation, without
any quantitative assessment of its validity.
Below we show that we instead observe a mean correlation length 
limited to a given fraction of the whole fault, 
and a rich dynamical regime leading to complex patterns of
possible slipping areas. 
The domains where avalanches can occur are not always maximal.
Therefore, it is clear that in this model it is not possible to have
a large earthquake at all times.
We will come back to this point in the Section ``Discussion''.
The next section contains the description of the model, while 
the numerical results are shown in section~\ref{sec:results}.

\section{Model}\label{sec:model}
The following model describes a one-dimensional fault with
$L$ units and with periodic boundary conditions.
Each unit $i$ represents the displacement $h_i$ 
of a plate with respect to a second one.
Plates are sliding with respect to each other and thus the displacement $h_i$ 
corresponds to a slip accumulated with time. 
An external field $\sigma_i$ characterizes the
speed of the strain accumulation in the unit: 
At each time step a unit $i$, chosen with probability 
$p_i\sim \exp(\beta \sigma_i)$, slips:
\beq\label{eq:add}
h_i \to h_i+1\;. 
\eeq
If $h_i$ forms a high gradient with one of its neighbors $j$,
in our case $h_i - h_j\ge 4$, a local elastic instability occurs.
This is relaxed by allowing the two nearest-neighbor units to get closer, 
\beq\label{eq:hihj}
h_i\to h_i-2\qquad \text{and}\qquad h_j\to h_j+2 \;.
\eeq
If this process leads to the formation of new unstable 
couples $(i,j)$, they are listed and processed into a 
random order until the list is empty, filling at the same time another list
with eventual new unstable pairs. 
The new list is then processed, and so on.
The iteration of this rule  leads to a final
state in which all bonds between units
are stable again. The whole avalanche of relaxations represents an earthquake
and is characterized by its
{\em size}  (the number of single relaxations, corresponding to the 
seismic moment), 
by its slipping {\em area} (the number of sites involved at least once), 
and by its epicenter (the unit where the avalanche started). 
It takes place by definition in one time step.
The waiting time between avalanches is then measured by the number of
time steps separating them. 

The aim of the field $\sigma_i$ is to reproduce some ``external'' 
tectonic loading, which should be originated by the  
crust portions that meet at the fault. 
Somewhat $\sigma$ replaces the loading calculated explicitly
with the laws of elasticity in other models (see for 
example~\citep{ben-zion96,ben-zion03:_intermittent}).
Since earthquakes play the main role in reshaping the stress field in the 
crust, we let each $\sigma_i$ evolve with a rule that couples it with the 
activity in the system: Every time that a redistribution 
(\ref{eq:hihj}) occurs, the two corresponding fields are set equal to their
average $\overline{\sigma}_{ij} = (\sigma_i+\sigma_j)/2$ plus a
noise term $\delta$ 
drawn at random (for each site) from the interval 
$[-1,1]$~\footnote{The choice of this interval just fixes the scale of 
fluctuations of the $\sigma_i$'s.}:
\beq
\label{eq:sisj}
\sigma_i  \to  \overline{\sigma}_{ij} +\delta_i
\qquad\text{and}\qquad
\sigma_j  \to  \overline{\sigma}_{ij} +\delta_j\;.
\eeq
The evolution of the system is thus stochastic in many aspects.
At the level of single redistributions involving (\ref{eq:hihj}) and 
(\ref{eq:sisj}), one has an update of $\sigma$'s with random $\delta$'s.
At the step (\ref{eq:add}) of forcing the system, the choice of $i$ according
to a probability $p_i$ is also stochastic. 
One can interpret the set of $\sigma_i$ as an array of local rates. Indeed,
a micro-slip ($h_i\to h_i+1$) takes place with a rate proportional to
$\exp(\beta \sigma_i)$.

A non-trivial regime emerges as long as $\beta$ is sufficiently large to 
lead to a persistence of the earthquake activity in areas of the system.
For $\beta\to\infty$  one finds a choice of the position 
to apply (\ref{eq:add}) that corresponds to the site with the largest $\sigma$.
This resembles an extremal dynamics for the field $\sigma_i$.
We rather chose $\beta$ large but finite, such that
many parts of the fault are likely to be active at the same time (if they
are share similar values of $\sigma_i$). The evolution of the
$\sigma_i$ guarantee a migration of active areas as well.

Despite the stochastic character of some of the microscopic updates, 
a rich phenomenology arises, with scale-free avalanches and 
with realistic interoccurrence statistics.

\section{Results}\label{sec:results}

{
We show results obtained by fixing $\beta=4$, 
which is large enough to lead to clustering of epicenters.
A preliminary check has shown qualitatively similar results in the 
range $2\le\beta\le 6$.
For each $L$, 
initial configurations for simplicity have $h_i =0$ and $\sigma_i =0$.
To be confident that the stationary regime has been reached,
we first run a long transient of 
$\approx 10^8\div 10^9$ time steps without collecting statistics.
From time step $t=0$ we then collect time series composed by
$2\div 3 \times 10^8$ time steps.
This constitutes a satisfactory statistics only if a large number of different 
profiles is sampled, which is the case for systems with  $\lesssim 2000$
units. We can thus collect data in
a reasonable time for systems up to this size.}

\begin{figure}[!tb]
\includegraphics[angle=0,width=8.0cm]{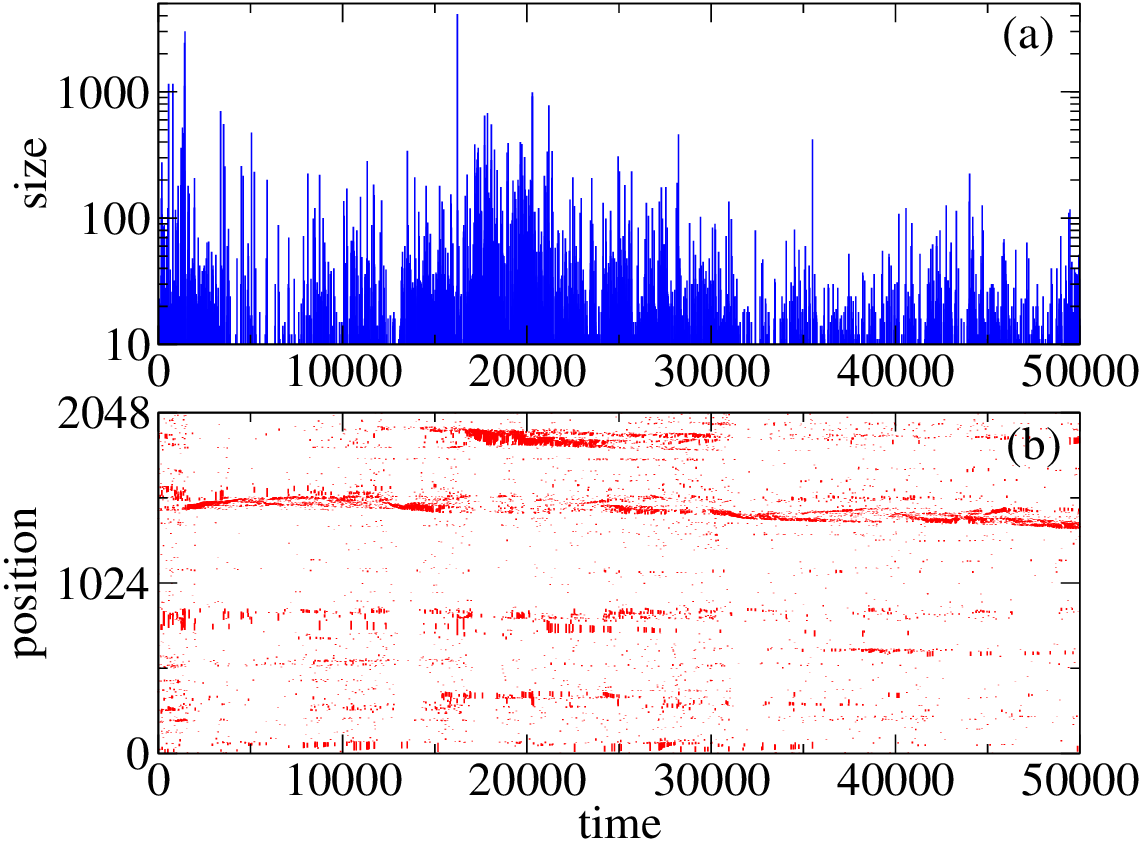}
\caption{Example of a time series for a system with $L=2048$ sites: 
(a) size vs time and (b) location of rupture areas versus time.
\label{fig:tser}
}
\vskip 0.3truecm
\includegraphics[angle=0,width=8.0cm]{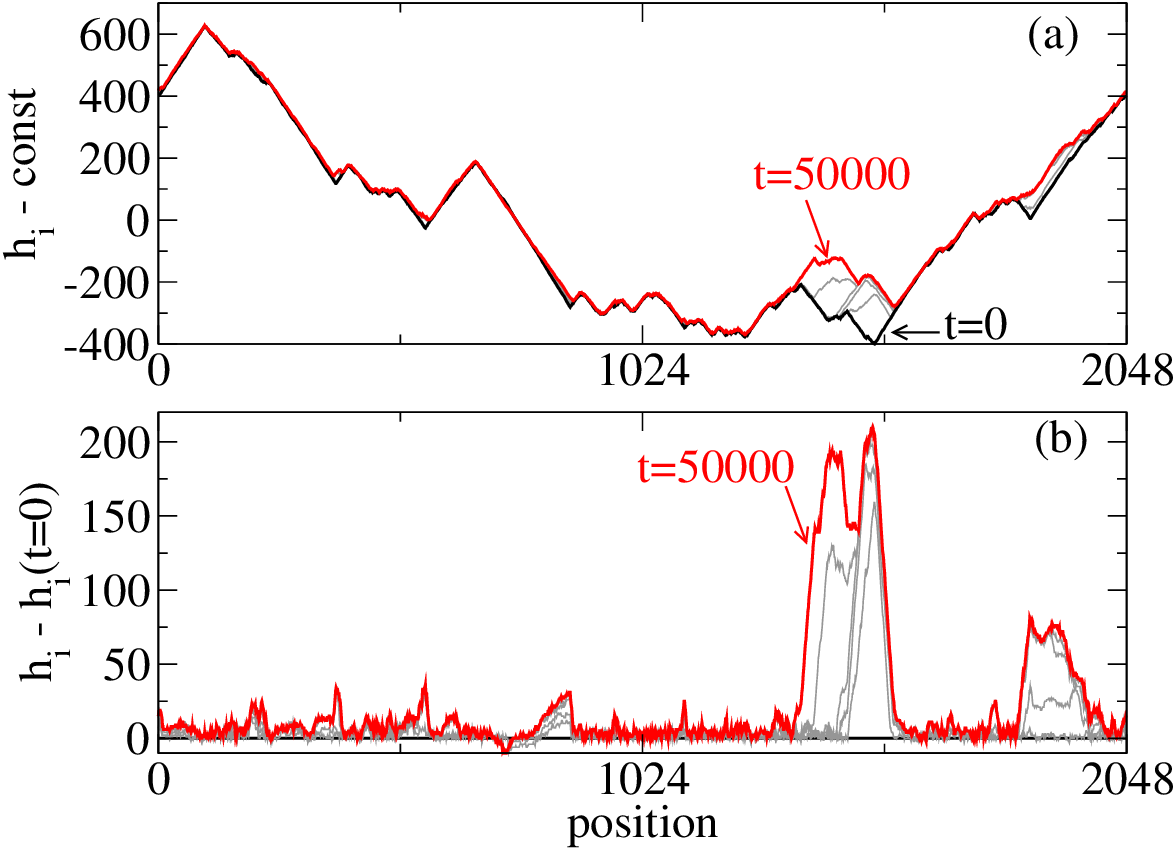}
\caption{(a) Profiles $h_i$ corresponding to the configuration at time
$t=0$ of Fig.~\ref{fig:tser} (black line) and at time $t=50000$ (red line),
and some intermediate stages (thin gray lines). To all curves we have subtracted
the average $h$ at time $t=0$.
(b) Difference of the same profiles with respect the initial one,
$h_i(t=0)$, to better visualize the regions where 
activity was concentrated in this example.
\label{fig:prof}
}
\end{figure}

A first glance at the behavior of the model is proposed in 
Fig.~\ref{fig:tser}, where we plot a sample of size and location of
rupture areas as a function of time. One can see that the activity is 
an alternation of earthquakes of several sizes, with a persistence in
active areas. This is confirmed by a plot of
the increment of $h_i$ with respect to the values at time
$t=0$: Fig.~\ref{fig:prof}(b) shows that the increments are 
concentrated in the active areas.

\begin{figure}[!tb]
\includegraphics[angle=0,width=8.0cm]{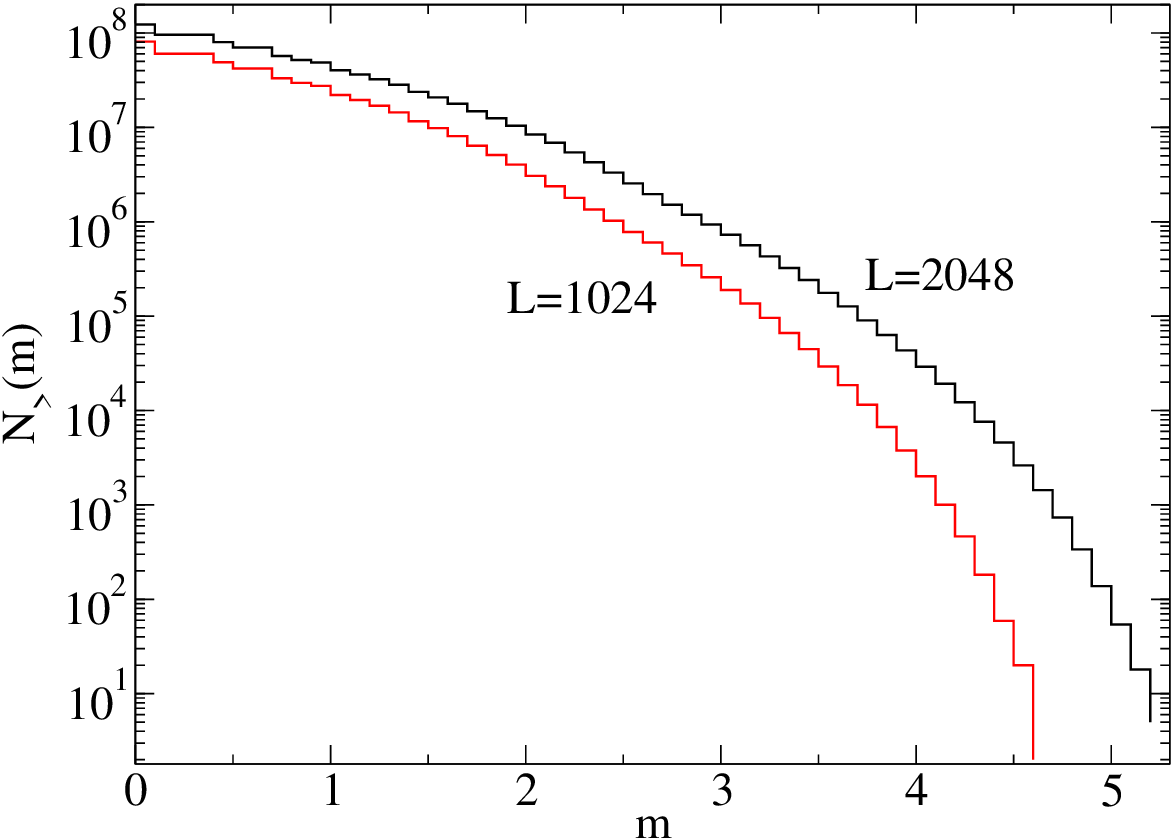}
\vskip 0.1truecm
\caption{Gutenberg-Richter law in systems with  $L=1024$ and 
$L=2048$.
\label{fig:Psize}
}
\end{figure}

\begin{figure}[!tb]
\includegraphics[angle=0,width=8.0cm]{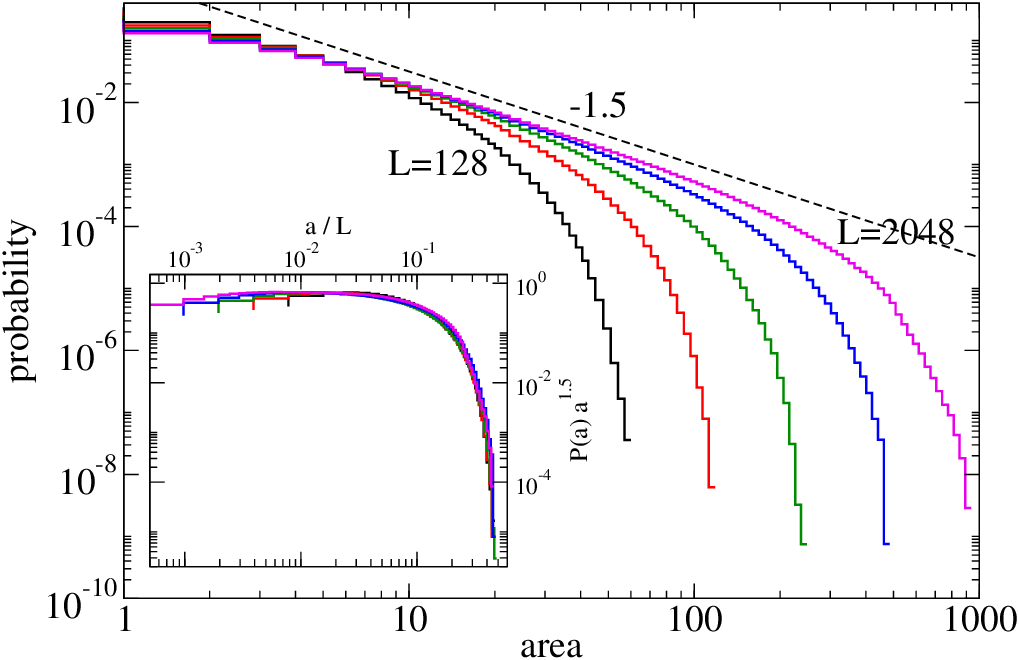}
\caption{Distributions of the area of avalanches, for 
$L=128$, $256$, $512$, $1024$, and $2048$.
Their power-law tail $\sim a^{-1.5}$ is highlighted by the dashed line. 
Inset: data collapse of $P(a) a^{\tau_a}$ vs $a/L$.
\label{fig:Parea}
}
\end{figure}

The statistics of several quantities turn out to be 
determined by power-laws. 
In order to display the frequency-size statistics,
we adopt the following definition of magnitude:
\[
m = \log_{10} s
\]
Note that the usual prefactor $2/3$~\citep{scholz_book} 
in the conversion from seismic moment to magnitude
is not suitable for a one-dimensional model because the area of events is
in fact a length.
In Fig.~\ref{fig:Psize} one can see that the number of events with
magnitude $\ge m$, denoted by $N_>(m)$,
{ seems to follow a GR law, $N_>(m)\sim 10^{-b m}$, with $b=1.1\pm 0.1$,
though  this distribution is most likely multiscaling, as it is often the case
in one-dimensional automata~\citep{kadanoff89:SOC}.
We postpone the exact characterization of this distribution to future work.
The distribution of slipping areas $a$ instead has a clearer scaling:
it develops a power-law tail 
 $\sim a^{-\tau_a}$ for increasing $L$, with $\tau_a=1.5$ (Fig~\ref{fig:Parea}),
and obeys to standard finite-size scaling
\begin{equation}
\label{eq:fssa}
P(a) \simeq a^{-\tau_a} F\left(\frac{a}{L^D}\right)
\end{equation}
with $D=1$ and where $F$ is a scaling function, 
see inset of Fig~\ref{fig:Parea}.
}

\begin{figure}[!tb]
\includegraphics[angle=0,width=8.0cm]{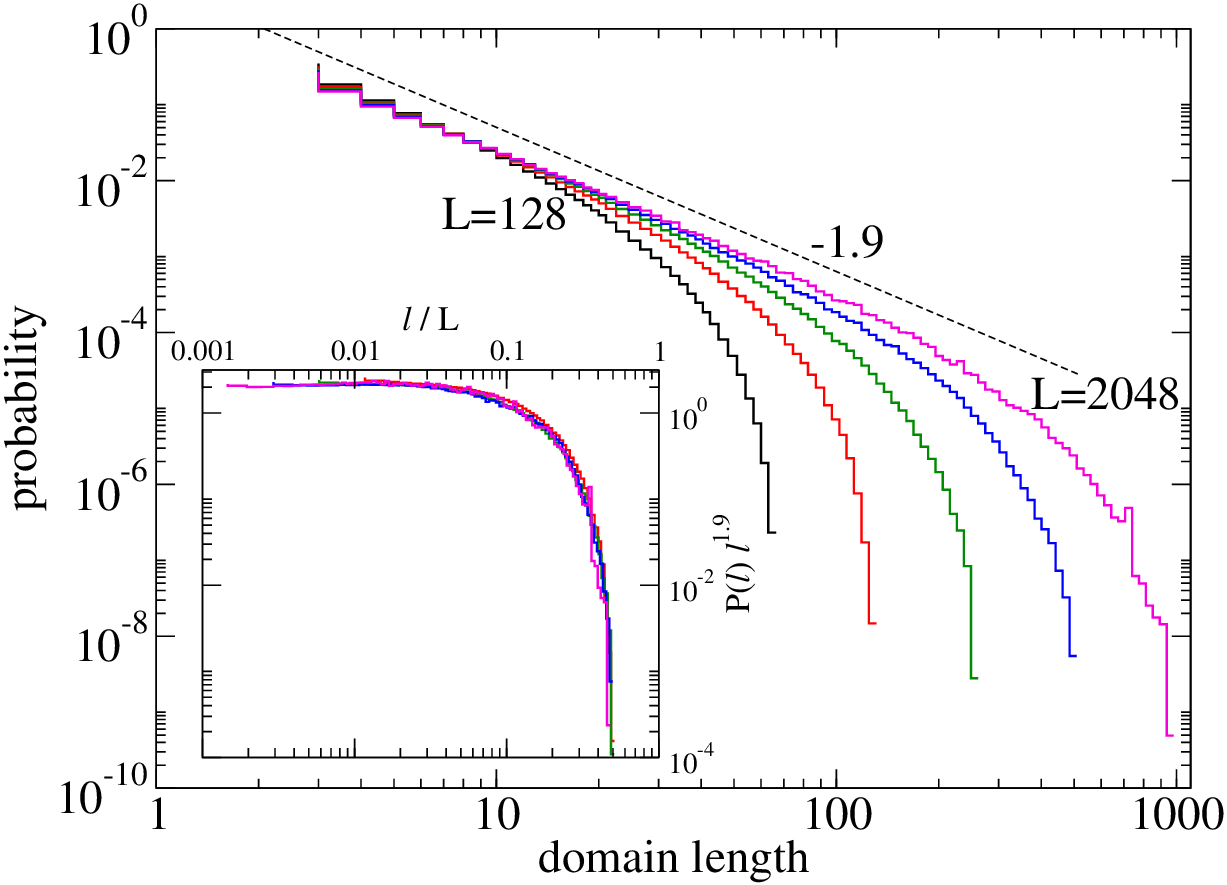}
\caption{Distribution of domain lengths $\ell$ 
(same $L$'s of Fig.~\ref{fig:Parea}).
The dashed line represents a power-law $\ell^{-1.9}$.
Inset: data collapse of $P(\ell) \ell^{\tau_\ell}$ vs $\ell/L$.
Data for the shortest $L=128$ are not included in the collapse.
\label{fig:Pdomain}
}
\end{figure}

In addition to the avalanche size and area, in this model we can 
also measure metric properties { characterizing the state of
the system between two avalanches}:
one is the length of {\em domains} of units having
constant sign in the slope of $h_i$.
Each profile $h_i$ is indeed an alternation of domains with increasing $h$
and domains of decreasing $h$, forming in general a non-trivial landscape,
see Fig.~\ref{fig:prof}(a).
This is also a result of the self-organization of the process,
which includes the evolution of the $\sigma_i$. 
{
Also domain lengths $\ell$ have a power-law distribution $\sim \ell^{-\tau_\ell}$
with $\tau_\ell\simeq 1.9$, see
Fig.~\ref{fig:Pdomain}, which displays finite-size scaling
\begin{equation}
P(\ell) \simeq \ell^{-\tau_\ell} G\left(\frac{\ell}{L^D}\right)
\end{equation}
also with  $D=1$ (inset of Fig.~\ref{fig:Pdomain}).
}
Since $\tau_a<\tau_\ell$, there is more chance to observe large areas than
large domains. On the other hand, avalanches take place within domains.
This suggests that avalanches are repetitive and appear more frequently
in long domains.

\begin{figure}[!tb]
\includegraphics[angle=0,width=8.0cm]{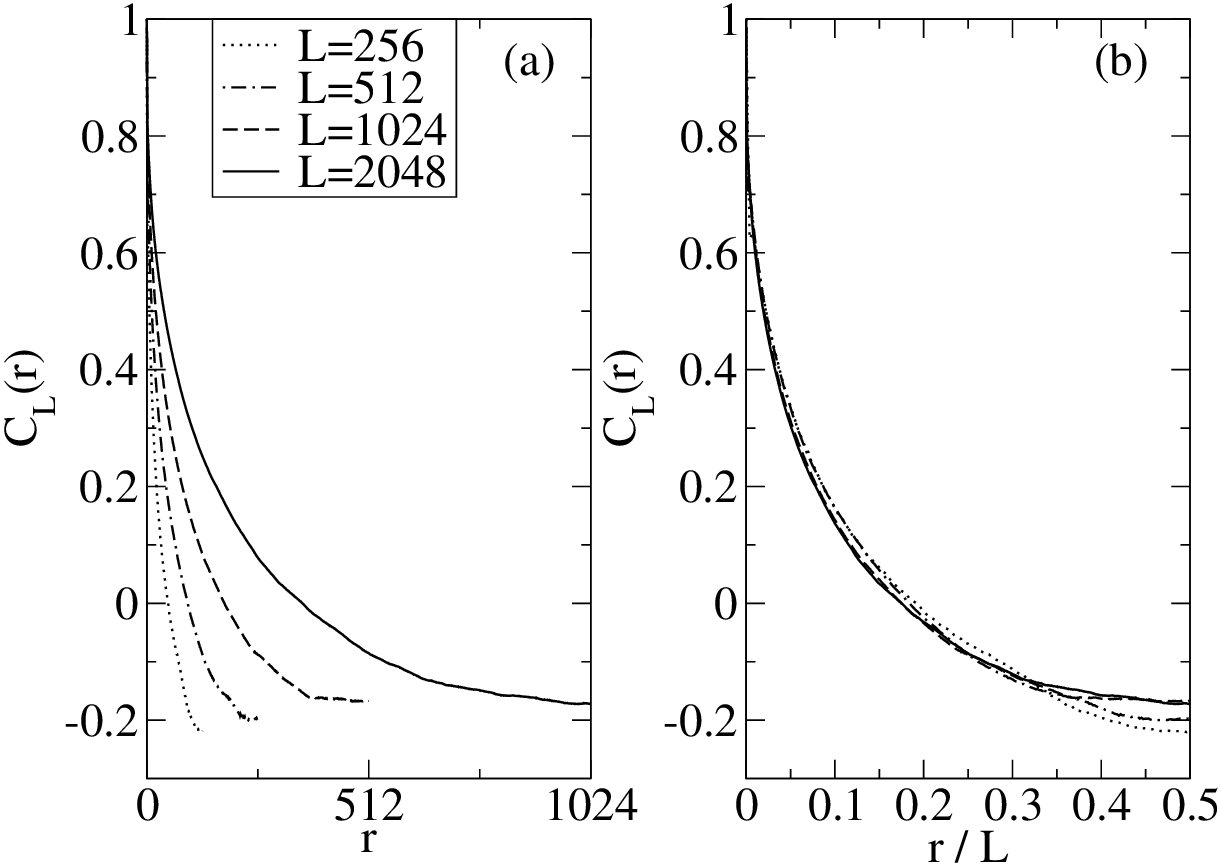}
\caption{Correlation function $C_L(r)$ of the stress $f_i$ 
for $L=256$, $512$, $1024$, and $2048$, plotted as a function of (a) $r$
and (b) $r/L$.
\label{fig:Corr}
}
\end{figure}

Connected with the scale-invariance of domains, there is also a scaling of the 
correlation length of the stress $f_i=h_{i+1}-h_i$ with the system size.
The correlation length can be read from the shape of the correlation function
\begin{equation}
C_L(r) = 
\frac{\mean{f_{i+r} f_{i}} - \mean{f_i}^2}{\mean{f_{i} f_{i}} - \mean{f_i}^2} 
=  \frac{\mean{f_{i+r} f_{i}}}{\mean{f_{i} f_{i}}}
\end{equation}
where $\mean{\ldots}$ means a statistical average over the sites and 
configurations\footnote{The periodic boundary conditions imply 
$\mean{f_i}=\mean{f_{i+r}}=0$.}.
It turns out that $C_L(r)$ conforms to a scaling
function $C_L(r)\simeq {\cal C} (r/L)$, with ${\cal C}(\ldots)$ 
independent on $L$, 
as shown in Fig.~\ref{fig:Corr}.
Hence, if we define the correlation length as the range where 
$C_L(r)>0.1$, we see (Fig.~\ref{fig:Corr})
that it has a value $\approx 10\% L$
that diverges linearly with $L$, 
as one expects in critical systems. 
We will come back to this point is the Discussion.

\begin{figure}[!tb]
\includegraphics[angle=0,width=8.0cm]{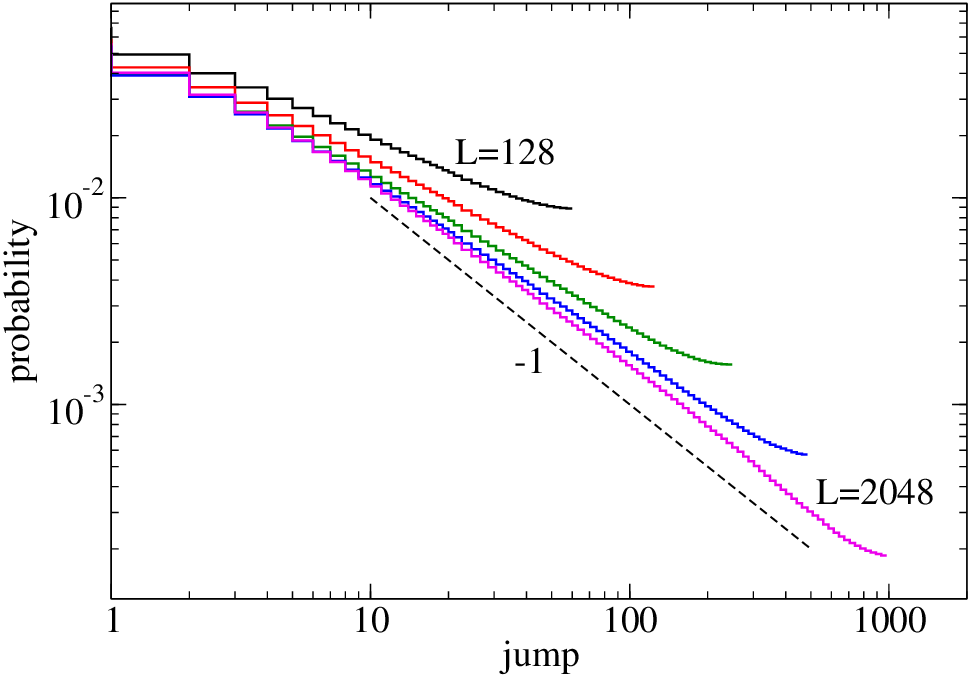}
\caption{Distribution of the jumps (distances between subsequent activities)
for the same  $L$'s of Fig.~\ref{fig:Parea}.
Their power-law tails have an exponent converging 
roughly to $\approx 1$ for large $L$.
\label{fig:Pj}
}
\end{figure}

Another quantity of interest is the {\em jump} 
between the position of grain addition at time
$t$ and the subsequent position of grain addition at $t+1$.
The jump distributions have also power-law tails,
with exponent converging to $\approx -1$, see Fig.~\ref{fig:Pj}.
This distribution is thus similar to that of distances between subsequent 
earthquakes~\citep{davidsen-paczuski05:jumps,corral06:_jumps}.
Also the crossover to a background level for long jumps
takes place at a length that is a fixed fraction
the size of the catalogue~\citep{davidsen-paczuski05:jumps,corral06:_jumps}.

\subsection{Temporal correlations}
During the last years part of 
the scientific debate on earthquake correlations has
been focusing on the statistics of waiting times between events, see
\citep{BM_05_soc} for an overview.
An issue was whether SOC models can have avalanches correlated with each
other. Some models have waiting times between avalanches with an 
exponential distribution, suggesting that their
events are completely uncorrelated. Clearly this is an unwanted feature 
in models of earthquakes.
Recently~\citet{bak02:_unified} 
and~\citet{corral03:_unified,corral04:_clustering,corral05:_bayesian} 
have shown that 
waiting times have in general a non-trivial scaling form in their 
distributions.

\begin{figure}[!tb]
\includegraphics[angle=0,width=8.0cm]{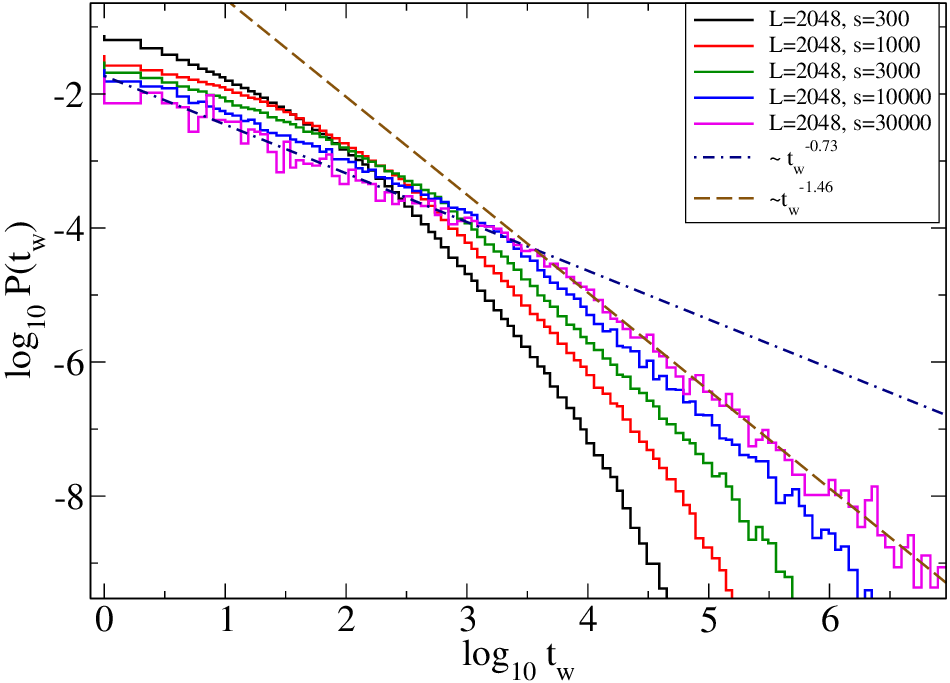}
\caption{Distribution of waiting times, for events larger than thresholds
$s$ ($L=2048$). 
Two power-law fits are also shown for the two parts of the distribution 
relative to $s=30000$.
\label{fig:Ptw}
}
\end{figure}

In Fig.~\ref{fig:Ptw} we plot some waiting time distributions that we observe
in our model, for $L=2048$ and for several minimum thresholds $s$ of the size.
These distributions have a shape with a double power-law form for high
thresholds, as observed in catalogs of regional seismicity
by~\citet{corral03:_unified}
and in an aftershock-sequence model by~\citet{lippiello07:ETAS}.

\begin{figure}[!tb]
\includegraphics[angle=0,width=8.0cm]{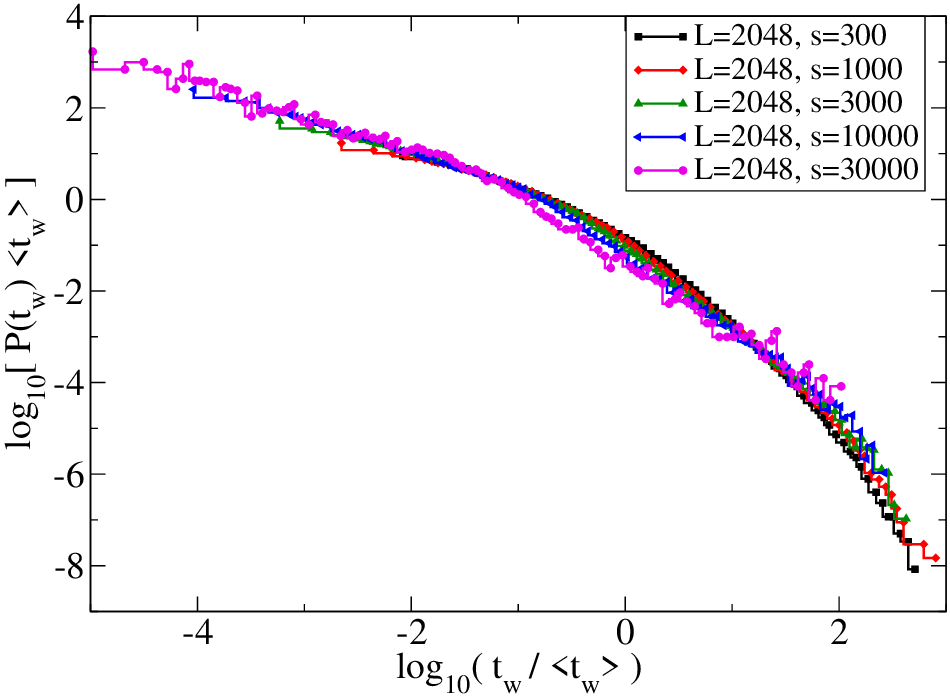}
\caption{Rescaled distribution of waiting times. $\langle t_w\rangle$ is
the mean waiting time between events (it depends on the threshold $s$).
\label{fig:Ptw_resc}
}
\end{figure}

In Fig.~\ref{fig:Ptw_resc} there is an attempt to collapse some of these
distributions on a single curve, by rescaling the waiting times to scales
in which their average value is $1$, that is, by multiplying their values by
the rate of events larger than the corresponding minimum thresholds.
This procedure revealed an interesting scaling form for real 
earthquakes~\citep{corral03:_unified,corral04:_clustering,corral05:_bayesian}
(and also for solar flares, see~\citep{BPS_05}):
in that case one observes a nice data collapse, with distributions 
being described by a single scaling function.
The data collapse for this model is only approximate.
We can conclude that the power-law tails in the distributions 
are a clear indication of a non-trivial organization and clustering 
in time of the avalanches, with some missing scale-invariance evidenced by
the thresholding procedure.

\begin{figure}[!tb]
\includegraphics[angle=0,width=8.0cm]{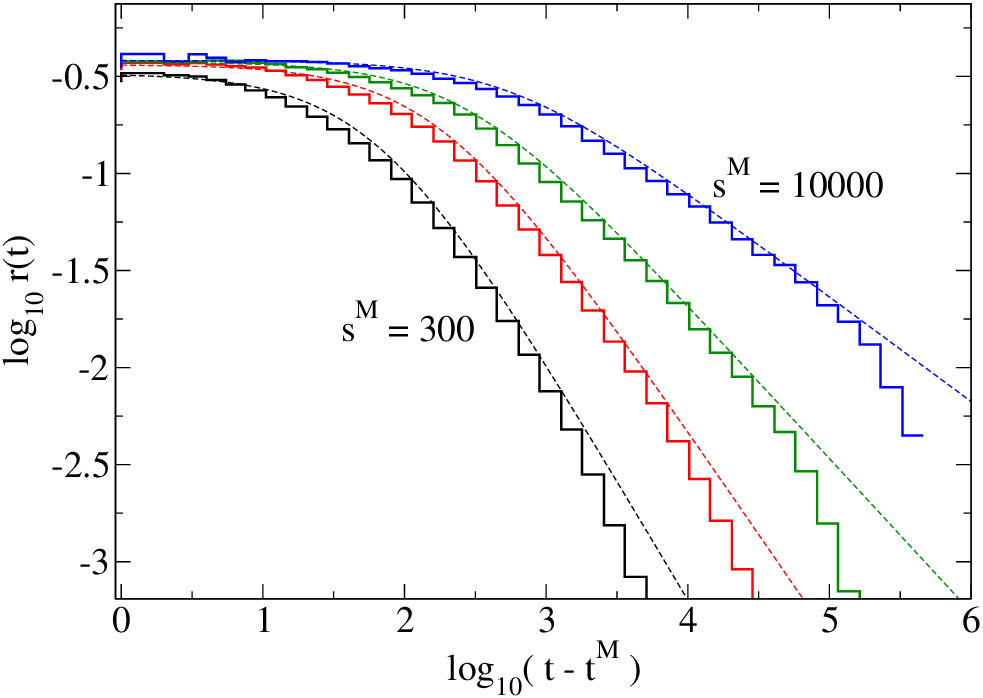}
\caption{Decay of aftershocks activity
after main shocks of size $s^M=300$, $1000$, $3000$, and $10000$,
in a system with $L=2048$.
Dense lines are data, while dashed lines are fit according to  
the generalized Omori decay (\ref{omori}).
\label{fig:omori}
}
\end{figure}

It is also not trivial to observe aftershocks in simple models of seismicity.
Indeed, one does not always observe Omori decay of aftershocks in synthetic
catalogs. However, this is a salient feature of seismicity,
characterizing the occurrence of correlated events even for 
years~\citep{utsu95:_omori,shcherbakov04:after,BP_04,BP_05,zaliapin08:after}.
{ Our model does not yield time series with patterns clearly identifiable
with aftershocks sequences, intended in the usual seismological sense.
Nevertheless, 
an Omori-like decay can be detected, confirming the temporal clustering
evidenced by waiting time statistics.
To visualize the Omori decay,}
we use a simple definition of aftershocks, leaving more
complicated spatio-temporal 
analysis~\citep{shcherbakov04:after,BP_04,BP_05,baiesi06,zaliapin08:after} 
for future works.
Let us consider events with size $s^M$ as main shocks 
(to improve the statistics, we actually consider events in a range
$[0.9 s^M,1.1 s^M]$).
Each of these events collects aftershocks in a time-window 
following its occurrence time $t^M$ and including only 
events of smaller size. 
This time window $t-t^M$ thus ends
if a new event of size at least $0.9 s^M$ occurs.
The averaged statistics of the rate $r(t-t^M)$
of avalanches after an main event of size $s^M$ 
is shown in Fig.~\ref{fig:omori} 
as a function of the time lag $t-t^M$ from the main 
shock, for several values of $s^M$.

One can see that the aftershock decays depend on $s^M$ and 
follow a generalized Omori decay
\begin{equation}\label{omori}
r(t)\sim \frac{A}{[1+(t-t^M)/t^*]^p}
\end{equation}
where A is a constant, $t^*$ is a characteristic time, and $p$ is the exponent
of the generalized decay (usually one observes $p\approx 1$).
As in real seismicity~\citep{BP_04,BP_05}, 
the onset of the power-law decay takes place at times  $t^*$
that increase with the size of the main event.
{ The same is true for the end of the Omori decay:
data in Fig.~\ref{fig:omori} have an exponential decay after the Omori regime,
as it was  found for aftershocks~\citep{BP_05}.
}
The exponent $p$ takes values ranging from $\approx 1.3$ for $s^M=300$,
to $\approx 0.5$ for $s^M=10000$.
Its variability somewhat reflects the same 
lack of invariance for increasing thresholds manifested by waiting-time 
distributions.

\section{Discussion}\label{sec:discussion}

Some previous SOC models with realistic phenomenology
are based on the mechanism of extremal 
dynamics~\citep{olami92:_OFC,hainzl99:SOC,hainzl00:SOC,
hergarten02:_OFC_aftershock,zoeller05:creeping,
lippiello05:_SOC_memory}, 
in which an earthquake starts always from the weakest unit.
Our stochastic model shows a more general mechanisms 
giving rise to correlated events within SOC, which
involves activity suitably
clustered in space and time, together with scale-free redistributions
of energy in the form of avalanches.
The random aspect cannot be excessive\footnote{
In our model, the activity spreads randomly in space with low $\beta$ values.
In this limit, domains shrink to exponentially short regions and 
the system loses scale-free avalanches.}:
a load completely random in space has been for years the standard 
in several SOC cellular automata, maybe because it is the simplest protocol. 
In the field of seismicity this choice 
is not supported by phenomenological observations, as we know that epicenters
are correlated and clustered.
When a random load was imposed, avalanches were found to be 
uncorrelated~\citep{BM_05_soc}.
We thus argue that a (correct) clustering in space of events cannot be
disentangled from the temporal clustering of events, both aspects being part
of the same global organization in critical systems.

Regardless of the lack of dissipation from
open boundaries, our process reaches a stationary critical regime. 
The reason is that its loading is not homogeneous and
the evolution via avalanches
generates the domains over which further large avalanches can occur.  
In the periodic system we have described, the minima of the 
accumulated slip profile are places where eventually avalanches must stop. 
These minima are not fixed but dynamic.

It is important to note that
the dynamics of the accumulated slip profile, with domains that evolve in time,
has non-trivial consequences. Each domain seems
to represent what is normally observed in canonical SOC systems with open
boundaries~\citep{bak_book}, the so called ``sandpiles'',
which have a profile with a single slope, from the maximum at
a closed boundary to a minimum at an open (dissipating) boundary.
Eventually the whole process somewhat resembles a collection 
of smaller homogeneous SOC systems, whose number and position fluctuates
in time. For each configuration, the
maximum correlation length should be close to the length of the longest domain.
Interestingly, this domain length is not always close to its possible maximum, 
which means that the system is often in a state incompatible 
with an earthquake spanning the whole fault. 
Moreover, we have seen that the range of the average correlation length
is a fraction of the system size. On the one side, this says that we have to
reconsider the typical value of correlation ranges upon change of scale
of the whole system.
Provided that we can meaningfully isolate an area from the
rest of the crust, on the other hand, we can expect a finite mean
correlation length within it.

Hence, our model does not reproduce a popular picture associated with SOC,
invoking a continuous state of ``maximal'' criticality in the crust 
due to an eventual infinite correlation length \citep{nature-debate}.
According to this picture,
earthquakes are inherently unpredictable in size, space and time because
their cascade to large events depends on minor details of the stress field.
This point has been used, for example, by~\citet{geller97:_science} 
to infer that earthquakes cannot be predicted.
The validity of their argument
can be limited by the lack of discussion about non-minor details.
These major details in our models are those that are macroscopically visible
when looking at the profile of the slip field $h_i$, namely the different
domains.
Unfortunately patterns like these are not accessible in real measurements.
Bak pointed out \citep{nature-debate} that an earthquake 
does not ``know how large it will become''.
This is not incompatible with our point that 
an earthquake ``knows how large it cannot become''. 
Perhaps both aspects should be taken into account in studies on earthquake 
prediction~\citep{keilis02:_rev}.

Therefore, according to our results, the following scenario is possible:
The process of self-organization in seismicity, due to the slow load of 
the crust and its fast relaxation via earthquakes, 
converges to a dynamical SOC regime, with rise and fall of patterns
of strongly correlated stress. These patterns may be associated with (local) 
fluctuating correlation lengths.

One could also have coexistence of SOC and 
other mechanisms~\citep{sammis-sornette02:_PNAS}.
A previous SOC model with a heterogeneous fixed pattern of faults 
\citep{huang98:_BTW-hier}
has a behavior consistent with the hypothesis that the approach to large
earthquakes is described by a critical-point 
picture~\citep{sykes_jaume99:_review,sammis-sornette02:_PNAS}, 
with a finite-time singularity of Benioff strain 
release and a divergence of a correlation 
length~\citep{zoeller02:_corr-length,zaliapin02:_corr-length}.
We have not investigated this point in our model yet,
though it seems that its dynamics does not break all the correlations
after a large earthquake. Indeed,
a large slip along a domain lowers the total energy stored in the system, 
and eventually shifts the domain range  of some units,
but the domain itself should be ready for 
similar earthquakes without too much effort.
However,
an eventual merging with other coherent domains might lead to an increase
of the correlation length in the area, with a possible connection with
previous studies~\citep{sykes_jaume99:_review,sammis-sornette02:_PNAS,zoeller02:_corr-length,zaliapin02:_corr-length}.
In any case, the stationary regime of our model appears to be different 
from that of {\em intermittent criticality}~\citep{ben-zion03:_intermittent,
bowman04:_intermittent}, in which every large event drives the system far from
criticality, which is then slowly restored by the dynamics.

\section{Conclusions}
We have shown that it is possible to build stochastic
processes with self-organized criticality
that reproduce several power-laws found in earthquake 
statistics, like the GR law, the generalized Omori laws,
the waiting-time distributions, 
and the distributions of distances between subsequent events.
The robust scale-invariant statistics  
generated by avalanches is the leading principle of the study.
We have stressed that it is important that the process generates 
activity clustered in space for eventually obtaining 
a clustering in time of events. { In our model, contrary to previous 
examples, the clustering takes place even if the system is stochastic, showing
that a moderate degree of randomness can be tolerated in SOC models with
spatio-temporal correlation.}

{ Our findings are contrary 
to a constant complete unpredictability of event sizes, even
if SOC is one of the main
mechanisms acting to generate the complexity of seismicity.
The point is that every SOC system can have a finite size.
We have described a system displaying domains that
resemble a fluctuating collection of canonical SOC cellular
automata. Interestingly, domains limit each other and 
their boundaries constitute the points where avalanches eventually
must stop. The size of each domain quantifies locally the correlation length,
which is thus a quantity fluctuating in space and time.
As a result, the mean correlation
length diverges with the system size, as expected, but it occupies only
a finite fraction of the system. 
We have been able to visualize these features thanks to the simplicity
of the one-dimensional model.}

The oversimplified process that we have discussed
is inspired by the phenomenology of earthquakes
and tries to encapsulate it,
but clearly it is a geophysical model in an embryonic stage. 
Hopefully the results and discussion we have presented 
provide new ideas that will be useful
for building models grounded on laws of geophysics and elasticity 
of solids, which still preserve the ability
to reproduce earthquakes phenomenology.
{
With models of this kind,
for example, it would be interesting to see if creeping sections of faults can
play the role of domain boundaries in the sense discussed in this paper.}

\begin{acknowledgements}
This research was supported by grant OT/07/034A from K.~U.~Leuven.
The author acknowledges discussions with
C. Maes and M. Paczuski, and warmly thanks J. Davidsen for the 
useful discussions as well as for the precious comments on the manuscript.
\end{acknowledgements}


\end{document}